# A local interstellar spectrum for galactic electrons

M. S. Potgieter*, R. R. Nndanganeni

*Centre for Space Research, North-West University, 2520 Potchefstroom, South Africa*

**Abstract**

A heliopause spectrum at 122 AU from the Sun is presented for galactic electrons over an energy range from 1 MeV to 50 GeV that can be considered the lowest possible local interstellar spectrum (LIS). The focus of this work is on the spectral shape of the LIS below ~1.0 GeV. The study is done by using a comprehensive numerical model for solar modulation in comparison with Voyager 1 observations at ~112 AU from the Sun and PAMELA data at Earth. Below ~1.0 GeV, this LIS exhibits a power law with $E^{-(1.55 \pm 0.05)}$, where $E$ is the kinetic energy of these electrons. However, reproducing the PAMELA electron spectrum averaged for 2009, requires a LIS with a different power law of the form $E^{-(3.15 \pm 0.05)}$ above ~5 GeV. Combining the two power laws with a smooth transition from low to high energies yields a LIS over the full energy range that is relevant and applicable to the modulation of cosmic ray electrons in the heliosphere. The break occurs between ~800 MeV and ~2 GeV as a characteristic feature of this LIS. The power-law form below ~1.0 GeV produces a challenge to the origin of these low energy galactic electrons. On the other hand, the results of this study can be used as a gauge for astrophysical modeling of the local interstellar spectrum for electrons.

*Keywords:* Galactic spectra; Cosmic rays; Galactic electrons; Heliosphere; Solar modulation; Heliosheath; Local interstellar spectrum

*Corresponding author: Address: Tel.: +27 18 299 2406; fax: +27 18 299 2421.
*E-mail address:* Marius.Potgieter@nwu.ac.za. (M.S. Potgieter)

1. **Introduction**

A crucially important aspect of the modulation of galactic cosmic rays in the heliosphere is that the local interstellar spectra (LIS) need to be specified as input spectra at an assumed modulation boundary and then modulated throughout the heliosphere as a function of position, energy and time. Because of solar modulation and the fact that the nature of the heliospheric diffusion coefficients is not yet fully established, all cosmic ray LIS at low kinetic energies ($E <$ ~1 GeV) remain controversial. This is true from an astrophysical and heliospheric point of view.

We focus on galactic electrons and approach the controversy from a heliospheric point of view. This is achievable because solar modulation models have reached a relatively high level of sophistication, including a third order tensor, supported by progress on diffusion and turbulence theory for the heliosphere. Even more important is that the Voyager 1 spacecraft is about to exit the inner heliosheath, the region between the solar wind termination shock (TS) and the heliopause (HP), while observing electrons between 6 and ~120 MeV [1]. Together with electron observations at Earth from PAMELA with $E >$ ~100 MeV [2], it is now possible to establish the total modulation between the heliospheric boundary and Earth, which provides a rather robust set of modulation parameters for comprehensive numerical modeling of solar modulation.

We first discuss the enduring controversy around electron galactic spectra (GS), focusing on $E <$ ~1 GeV, and the conventional assumption that these GS are LIS. The solutions of a numerical model for the modulation of galactic electrons in the heliosphere are then presented in comparison with Voyager 1 and PAMELA observations to obtain a computed electron spectrum at the heliopause that can be considered the lowest possible very LIS.

2. **Galactic spectra assumed as local interstellar spectra**

Computed galactic cosmic ray spectra, from a solar modulation point of view, are referred to as spectra that are produced from astrophysical sources, usually assumed to be evenly distributed through the Galaxy, typically very far from the heliosphere. More elaborate approaches to the distribution of sources have also been followed (e.g., [3]), even considering contributions of sources or regions relatively closer to the heliosphere (e.g. [4]) and with new theoretical (e.g. [5]) and observational developments [6]. These spectra are calculated using various approaches based on different assumptions, but mostly using numerical models, e.g. the well-known GALPROP propagation model [6,7,8]. For energies below several GeV, which are of great interest to solar modulation studies, the galactic propagation processes are acknowledged as to be less precise as illustrated comprehensively by [9]. The situation for galactic electrons at low energies has always been considered somewhat better because electrons produce synchrotron radiation, so that radio data assist in estimating the electron GS at these low energies. For a discussion of this particular approach and some examples of consequent electron GS, see Langner et al. [10], Strong et al. [6] and Webber and Higbie [11]

Computed galactic spectra usually do not contain the contributions of any specific (local) sources within parsecs from the heliosphere so that an interstellar spectrum may be different from an average GS (typically as is computed with GALPROP). An averaged interstellar spectrum may be different from a LIS (thousands of AU away from the Sun), which might be different from a very LIS, say within ~200 AU away from the Sun, or what can be called a heliopause spectrum, right at the edge of the heliosphere. If known, the latter would be the ideal spectrum to use as an input spectrum for solar modulation models. This study is used to compute such an electron spectrum from a heliospheric point of view, using an existing comprehensive modulation model and the mentioned observations.

Potgieter and Ferreira [12] and Potgieter and Langner [13] showed that the heliospheric TS could in principle re-accelerate low-energy galactic electrons to energies as high as ~1 GeV so that a heliopause spectrum could be different from a TS spectrum. Principally, such a TS spectrum may even be higher than a GS spectrum, depending on the energies considered. However, because the TS was observed as rather weak [14], only a factor of ~2 increase was observed close to the TS for 6-14 MeV electrons. Since the crossing of the TS, Voyager 1 has observed an increase of a factor of ~ 60 on its way to the HP [1]. During the period 2009 to 2012, the intensity of these low-energy electrons has increased by almost a factor of 10. This means that the HP spectrum is at least a factor of ~30 higher at these energies than what had been observed by Voyager 1 at the TS, more than 8 years ago. It is thus reasonable to neglect the influence of the heliospheric TS on a very LIS when cosmic ray ions are considered, but the possibility of the re-acceleration of low energy electrons (and positrons) inside the heliosphere will have to be investigated before a decisive conclusion can be made. The latter is not addressed in this work. For a review on the solar modulation of cosmic rays see [15].

For a compilation of computed GS based on the GALPROP code for cosmic ray protons, anti-protons, electrons, positrons, Helium, Boron and Carbon and more, see Moskalenko et al. [7] and Ptuskin et al. [9]. For an update on electron LIS from an astrophysical point of view, see Strong et al. [6]. For a discussion on the contribution of Jovian electrons to the global solar modulation of electrons, see Strauss et al. [16] and Potgieter and Nndanganeni [17]. For the rest of this report, the focus remains strictly on galactic electrons.

Figure 1 is presented as an example of the mentioned uncertainties in GS that were used as LIS, from a historical point of view. It shows several electron spectra as computed with earlier versions of the GALPROP propagation model, done between 1994 and 2000. The highest electron GS was reported by Strong et al. [18] and was used frequently as the LIS in earlier modulation studies (e.g. [21,22]). Later, Strong et al. [19] argued that this GS was too high and much lower spectra were computed as indicated in the Figure. Langner et al. [10] recalculated the electron GS using a phenomenological approach, including also radio data for calculations at lower energies. They found a spectrum that was between the spectra reported by [18] and [19]. This spectrum is shown together with the GALPROP computations in units of particles $m^{-2}\ s^{-1}\ sr^{-1}\ MeV^{-1}$. A feature of these GS is that all of them deviate from a power law progressively below ~1 GeV to differ substantially the lower the energy becomes. A power-law form is recognizable above ~1 GeV but at lower energies this is not the case. Evidently, the situation has been unsatisfactorily, yielding a difference of a factor of 1000 at 1 MeV. It also emphasizes that GS for electrons at these low energies have remain controversial over a long period of time.

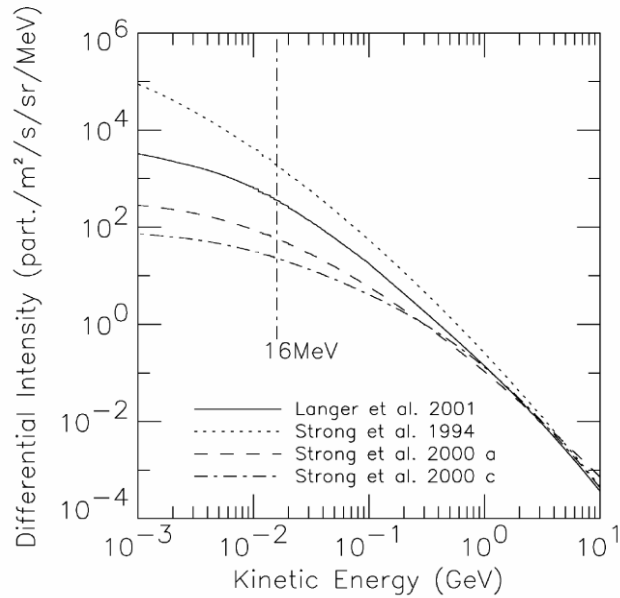

**Figure 1:** Computed galactic spectra for low-energy electrons by Strong et al. [18,19] are shown in comparison with the spectra by Langner et al. [10]. Figure is adapted from Ferreira et al. [20] which focused on the solar modulation of 16 MeV electrons in the heliosphere.

Ptuskin et al. [9] investigated and re-examined some of the physical processes in galactic space, involving in particular the rigidity dependence of the diffusion of cosmic rays through the Galaxy. They applied three different theoretical approaches in the GALPROP code: diffusive reacceleration with damping (called the DRD model); plain diffusion (PD model) with an ad hoc break in the galactic diffusion coefficient and an approach with distributed reacceleration (DR model) and power law diffusion with no breaks. The PD and DR models give a GS with relatively high intensities at low energies, while the DRD model produces a completely different GS, with much lower intensities for $E < \sim 3$ GeV. They also presented modulated spectra at Earth obtained by using a rudimentary force-field modulation model (see the review by Quenby [23]) in comparison with observations. This modulation model is not valid for electron modulation at $E < \sim 1$ GeV because it handles adiabatic energy losses as if electrons are losing as much energy as protons, which is not the case. For a comprehensive illustration of electron modulation including particle drifts, see [21,28].

Webber and Higbie [11] followed up on the work of Ptuskin et al. [9] and used a Monte Carlo Diffusion Model for electron propagation in the Galaxy to calculate alternative electron GS below ~1 GeV. They referred to these GS as interstellar spectra. This was the first attempt to actually compute a LIS (to be compared to Voyager 1 data) instead of simply assuming the GS to be the LIS. They also emphasized that solar modulation effects should be properly handled at these lower energies in the outer heliosphere to reconcile Voyager 1 observations with those at Earth when using computed GS (see also Langner et al. [10]). The force-field model cannot be used for observations in the outer heliosphere where adiabatic energy losses are almost negligible (see e.g., Caballero-Lopez and Moraal [24]). Webber and Higbie [11] calculated several plausible LIS and found most of them lower than the GS from Langner et al. [10]. They also critically evaluated the electron spectra presented by Ptuskin et al. [9] and found them either too high or too low, depending on the energy considered. Three of their spectra, identified as IS2.2, IS2.3 and IS2.4, are basically the same down to ~500 MeV with their IS2.3 spectrum very similar to the GS by [10]. Since both groups utilized observed radio data to improve the spectral shape of these spectra at energies $E < \sim 50$ MeV, it is not so surprising that their spectral forms are more or less similar. However, they computed different differential flux values at these very low energies. Unfortunately, the spectra from [11] were presented only up to 10 GeV, and their final spectra only down to 10 MeV.

Inspection of all the LIS that Webber and Higbie [11] computed or reported on, indicates that some of them are not consistent with electron observations in the heliosphere and can be eliminated. The peculiar forms of the PD and DRD approaches make them unsuitable as LIS for modulation studies because they cannot reproduce electron observations in the outer heliosphere. The obviously unsuitable one is the DRD spectrum because it touches the observations at Earth at ~ 0.7 GeV made during 1998, thus illustrating an inconsistency with respect to heliospheric observations. The Webber and Higbie spectra indicated as IS9 and IS11 are also too low at $E < 100$ MeV because the present Voyager 1 measurements are already surpassing this level. These spectral shapes can thus also be considered as improbable LIS at $E < 100$ MeV. The point we want to make is that consensus does not exist, quite evidently, about what the value or the energy dependence (spectral slope) of an electron LIS is below ~1 GeV, where solar modulation is significant. We assume that solar modulation becomes insignificant with $E > 30$ GeV (in this context, see Potgieter and Strauss [33]).

The prime objective of this modelling study is to use the available Voyager 1 observations of galactic electrons between 6 MeV to ~120 MeV at a distance beyond 112 AU from the Sun to determine a heliopause spectrum that can be presented as the lowest possible LIS for these low energy electrons. We consequently determine the spectral slope at these energies, and then compute the differential flux values between 1 MeV and 200 MeV. Unfortunately, the Voyager electron observations extend only up to ~ 100 MeV. The spectral slope is then adjusted with a different power law index above ~1 GeV to reproduce the modulated PAMELA electron data at Earth [2,25,26]. The PAMELA electrons unfortunately extend down to only ~200 MeV. This work follows up on the report by Potgieter and Nndanganeni [17] where the emphasis was on the total modulation of electrons in the heliosphere and to predict the intensity of low energy galactic electrons at Earth. Because electron observations at these low energies are seriously influenced by Jovian electrons, the true galactic intensity cannot be observed directly. This report is seen as complimentary to what was reported by [17].

The electron spectrum at or beyond 112 AU is an optimal choice from the available Voyager 1 data in order to determine a heliopause spectrum because earlier spectra, closer to the TS, are much lower and may be subjected to short-term changes because the region closer to the TS is more turbulent as it shifts position with changing solar activity [14].

## 3.  Numerical modulation model

A full three-dimensional (3D) numerical model, with three spatial dimensions and an energy dependence (four computational dimensions), was used to compute electron spectra at selected positions in the heliosphere, including the inner heliosheath. This model was described by [17] and only changes made are indicated here. In short, the model is based on the numerical solution of Parker's transport equation [25] for solar modulation

$$\frac{\partial f}{\partial t} = -\left(\mathbf{V} + \langle \mathbf{v}_d \rangle\right) \cdot \nabla f + \nabla \cdot \left(\mathbf{K}_s \cdot \nabla f\right) + \frac{1}{3}(\nabla \cdot \mathbf{V})\frac{\partial f}{\partial \ln P}. \quad (1)$$

Here, $f(r,P,t)$ is the cosmic ray distribution function, $t$ is time, $P$ is rigidity, $r$ is the position in 3D, with $r, \theta, \varphi$ the radial, polar and azimuthal coordinates in a heliocentric spherical coordinate system. We assumed $\partial f / \partial t = 0$ which means the contribution of short-term modulation effects is neglected with the focus instead on the global trends. Terms on the right hand side respectively represent convection, with $\mathbf{V}$ the solar wind velocity; gradient and curvature drifts, with $\langle \mathbf{v}_d \rangle$ the particle drift velocity; diffusion, and adiabatic energy changes. A full diffusion tensor $\mathbf{K}_s$ was used, as determined by the diffusion coefficients parallel and perpendicular to the average heliospheric magnetic field.

The convection process was changed across the TS to incorporate the much slower solar wind speed beyond the TS. The solar wind speed also changes its latitude dependence with solar activity.

We assumed the TS at 94 AU and the HP at 122 AU (noted that [17] used 120 AU) to be consistent with Voyager 1 observations. This means the heliosheath is 28 AU wide in the direction in which Voyager 1 is moving. It is reported by Webber and McDonald [26] that Voyager 1 crossed a region that looks like the HP in August 2012 but this needs to be confirmed because the corresponding magnetic field observations have not responded to what would be expected from a HP crossing.

The spatial dependence of the diffusion coefficients was specified differently over the first 80 AU (basically up to the TS region where the turbulence start to change significantly) than in the outer heliosphere. Their rigidity dependence was specified differently above and below 0.4 GV [17,21]. This is in accord with expectation that the turbulence at and beyond the TS (inside the inner heliosheath) is to increase with subsequent lower diffusion coefficients. The parallel and two perpendicular diffusion coefficients, in the radial and polar direction, and the drift coefficient were used as described and motivated by [17]. We focus on solar minimum modulation conditions.

## 4. Results and discussion

The process to reproduce the Voyager 1 electron data at ~112 AU consists of adjusting the input spectrum (what we call a heliopause spectrum), specified at the simulated HP for the 3D model until the compatibility to the mentioned observational data was optimal. Simultaneously, the modulated electron spectrum observed by PAMELA [27] at Earth for 2009 had to be reproduced with one set of modulation parameters. The LIS from Langner et al. [10] was used as point of departure. It was adjusted at higher energies (> 50 GeV) to reproduce the PAMELA spectrum. It was quickly realized that the various forms of the LIS as mentioned above were not suitable at lower energies. After a tedious process, the result as shown in Figure 2 was obtained. Here, the modulated spectrum at Earth is shown in comparison with the PAMELA spectrum at Earth and at 110 AU in comparison with the Voyager 1 data at ~112 AU, both with respect to the computed heliopause spectrum (the upper solid line) as required for reproducing simultaneously the two sets of observations. The comparison with the Voyager data indicates that in order to reproduce the spectral shape of the observed spectrum at ~112 AU, the LIS below ~1.0 GeV must have a power law form. Taking statistical and systematic uncertainties into account this form can be refined to be given as $E^{-(1.55 \pm 0.05)}$.

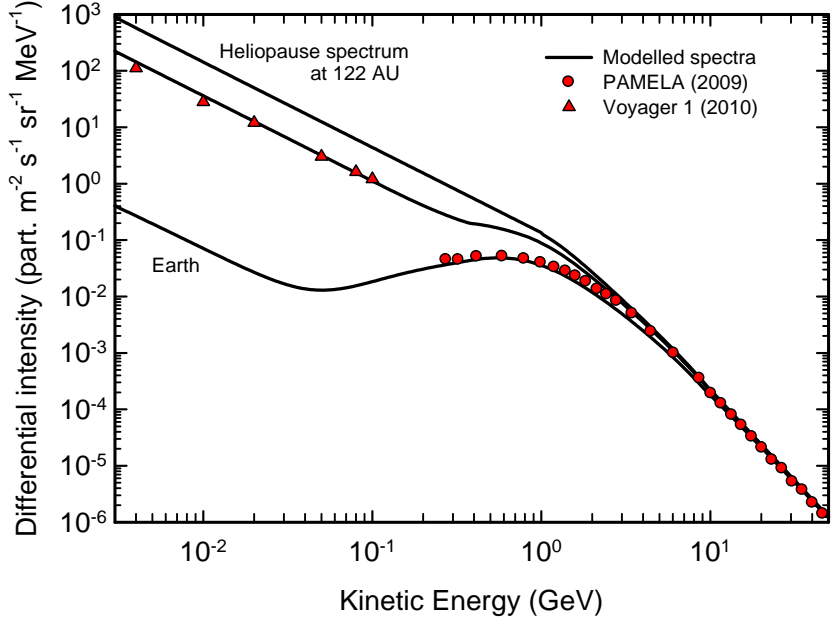

**Figure 2:** Computed modulated electron spectra at Earth (with polar angle of $\theta = 90°$) and at 110 AU (with $\theta = 60°$ corresponding to the Voyager 1 trajectory). These spectra are compatible with observations from Voyager 1 during 2010 at ~112 AU and from PAMELA at Earth during 2009 as indicated. The heliopause spectrum (upper solid line) is specified at the heliopause, positioned in the model at 122 AU.

Compatibility with the PAMELA spectrum requires that the LIS must have a power law form with $E^{-(3.15 \pm 0.05)}$ above ~ 5 GeV. This is consistent with the spectral index reported for 40-50 GeV by the PAMELA group [e.g., 2], who reported a power law of $E^{-(3.18 \pm 0.05)}$. Combining the two power laws with a smooth transition from high to low energies produces the upper solid line shown in Figure 2. The second important result is that the break between the power laws occurs between ~800 MeV and ~2 GeV. More on this aspect follows below. We consider this spectrum as the heliopause spectrum which is the lowest possible LIS for electrons.

As emphasized by [17], the heliospheric diffusion coefficients for electrons must be independent of energy at these low energies to reproduce the Voyager observations. If any weaker or strong rigidity dependence is used, the $E^{-1.55}$ spectral slope of the spectrum at 112 AU cannot be maintained and will not give the steady modulation (difference between the modulated spectrum and the LIS is unchanged) over an energy range from 1 MeV to ~ 200 MeV at 112 AU. This apparent independence is consistent with heliospheric turbulence theory as discussed by [17]. They also pointed out that the heliosheath acts as a very effective modulation 'hurdle' for these low energy electrons, causing intensity radial gradients of up to 20% $AU^{-1}$ as modelled by Nkosi et al. [31]. It should be kept in mind that the Voyager 1 detector cannot distinguish between electrons and positrons whereas the PAMELA detector can.

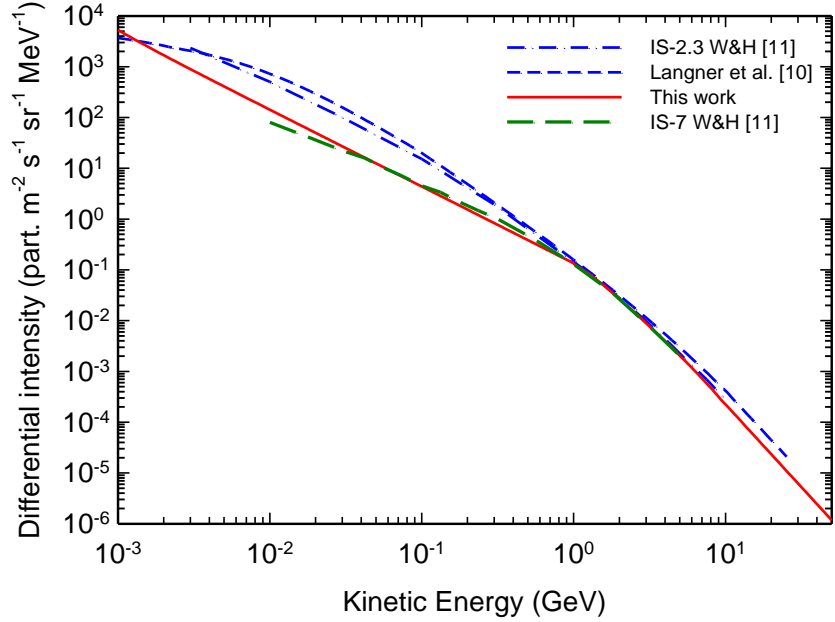

**Figure 3:** Computed electron heliopause spectrum (solid red line) compared to previously reported GS for which radio data were used to estimate the GS below ~100 MeV: Spectrum IS7 from Webber and Higbie [11] is the green long-dashed line; their spectrum IS2.3 is the dark red dash-dotted line and the Langner et al. [10] spectrum is the blue dashed line. This new LIS consists basically of two power-laws, changing from $E^{-3.15}$ at 40-50 GeV to $E^{-1.55}$ below ~ 0.8 GeV. The full expression allowing for a smooth transition is given by Eq. (2).

In Figure 3, the LIS presented in Figure 2 is compared to three other published LIS where Voyager observations were considered. They are IS2.3 and IS7 from Webber and Higbie [11] and the LIS from Langner et al. [10]. Both groups used radio data to estimate the GS at these low energies. A broader comparison was made between several computed LIS by [11] as discussed above and is not repeated here. It follows that the LIS presented here is lower than the LIS by [10] at almost all energies but the IS7 spectrum from [11] is quite similar at energies above 1 GeV although the latter stops at ~8 GeV. Below 1 GeV, the IS7 spectrum has a slightly lesser spectral index which causes it to become progressively lower with decreasing energy than what a single power law would give.

Next, a few comments are made on the break in the spectral shape of the presented HP spectrum. This was not addressed by [10,11,17]. It should be noted that the relation between synchrotron data and electrons is complicated by the presence of secondary electrons and positrons in the Galaxy so that it is not a straightforward matter to determine the spectral slope below a few GeV with galactic propagation modeling. Strong et al. [6] re-analysed synchrotron radiation data using various radio surveys to constrain the low-energy electron LIS in combination with data from Fermi-LAT and other experiments. After a comprehensive investigation, but still using a force-field modulation model, they concluded that the electron LIS must turn (exhibits a spectral break) below a few GeV in line with what is presented here. In addition they stated that the LIS below a few GeV has to be lower than what standard galactic propagation models predict, again consistent to what is shown here. They studied and presented several models which differ significantly below 100 GeV. Studying these different LIS (see their Figures 1, 4, 8, 11 and 12) indicate that the LIS presented in their Figure 12 resembles ours closely below ~800 MeV but not above this energy. Most of their models have more-or-less the same slope above 3 GeV (because they resemble trends in observations) but differ with ours in terms of absolute value. The model in the top left panel of their Figure 4 seems to be the most compatible with our result. A common factor as exhibited by all their spectra is the break in the power law. Our results should be useful to determine the optimal parameter set for a propagation code such as GALPROP.

Further investigation from an astrophysics point of view seems required to establish if the $E^{-1.55}$ spectral slope below a few GeV, as found here, has a galactic origin. Or, could it perhaps be influenced by processes inside the heliosphere? Concerning the latter issue, it is rather curious that if the TS was considered a strong, plain shock with a compression ratio of $s = 4$, the spectral shape of the consequent TS spectrum is $E^{-1}$, whereas for a weaker TS, with $s = 2.5$, it becomes $E^{-1.5}$. The TS is much more complicated than a plain shock, so that this aspect also requires further study from a heliospheric modulation point of view. The important point that we want to emphasize is that we found, with a solar modulation approach, a break between the mentioned two power laws in the LIS for electrons at kinetic energies very similar to the work based on an astrophysics approach.

A final comment is that cosmic ray modulation beyond the HP has become a very relevant topic since both Voyager spacecraft are about to explore the outer heliosheath (beyond the HP) and therefore may actually measure a pristine LIS sooner than what we anticipate. Scherer at al. [29] argued that a certain percentage of modulation may occur beyond the HP. Recently, Strauss et al. [30] followed this up and computed that the differential intensity of 100 MeV protons may decrease by ~25% from where the heliosphere is turbulently disturbed (inwards from the heliospheric bow wave up to the HP). However, because the diffusion coefficients of low energy electrons are independent of energy, making them significantly larger than for protons of the same rigidity, this percentage should be much less for 100 MeV electrons. Our computation for the LIS as specified at 122 AU from the Sun takes this into account. However, it could be that we underestimate this effect and that the true electron LIS is higher. It appears from electron counting rates (not differential flux) reported by [26] that Voyager 1 observed a significant increase over a relatively short period of time at what appears to be the heliopause. Future Voyager 1 observations should enlighten us in this respect as it moves outwards at 3 AU per year.

The expression for the LIS presented here, in terms of differential intensity in units of particles m$^{-2}$ s$^{-1}$ sr$^{-1}$ MeV$^{-1}$ and as a function of kinetic energy $E$ from 1 MeV to 50 GeV, is as follows:

If (1.0 MeV $\leq E \leq$ 1.0 GeV): $\quad j_{LIS} = \exp[-2.0 - 1.511(\ln E)]$,

if (1.0 GeV $< E \leq$ 10.0 GeV): $\quad j_{LIS} = \dfrac{0.1349 - (6.6 \times 10^{-3})E + (15.49 \times 10^{-5})E^2}{1 - (1.3187)E + (1.0810)E^2 + (0.2327)E^3}$, (2)

and if $E >$ 10.0 GeV : $\quad j_{LIS} = \exp\left[-0.89 - 3.262(\ln E)\right]$.

This heliopause spectrum gives the differential intensity for 10 MeV electrons as (145±10) electrons m$^{-2}$ s$^{-1}$ sr$^{-1}$ MeV$^{-1}$, considered to be the lowest possible LIS value at this energy. The most recent reported observation (in 2012) from Voyager 1 made between 6-14 MeV seems to have reached a level of at least 85 electrons m$^{-2}$ s$^{-1}$ sr$^{-1}$ MeV$^{-1}$, with a short-term time variation of at least 15 electrons m$^{-2}$ s$^{-1}$ sr$^{-1}$ MeV$^{-1}$ [see 32]. Unfortunately, an observed and updated spectrum from 6-100 MeV beyond the apparent HP is not yet available.

## 5. Conclusions

We present a heliopause spectrum for cosmic ray electrons over an energy range from 1 MeV to 50 GeV that can be considered the lowest possible LIS. This is done by using a comprehensive numerical model for solar modulation in comparison with Voyager 1 observations during 2010 at ~112 AU from the Sun and PAMELA data at Earth from 2009.

We report that below ~1.0 GeV this LIS has a power law form with $E^{-(1.55 \pm 0.05)}$. However, in order to reproduce the PAMELA electron spectrum for 2009, the LIS must have a different power law form with $E^{-(3.15 \pm 0.05)}$ above ~ 5 GeV. The latter is consistent with the spectral index reported for 40-50 GeV by the PAMELA group of $E^{-(3.18 \pm 0.05)}$. Combining the two power laws with a smooth transition from low to high energies produces the LIS shown as the solid line in Figure 3. The reported break between ~800 MeV and ~2 GeV in the spectral form of the LIS seems consistent with the results of Strong et al. [6] who studied the electron LIS from an astrophysics point of view.

The point that we want to emphasize is that we found a break between the mentioned two power laws in the LIS for electrons at kinetic energies very similar to the work based on an astrophysics approach. The results of this study can be used as a gauge for astrophysical modeling of the local interstellar spectrum for electrons and should be investigated further.

**6.     Acknowledgements**

The authors thank Bill Webber for providing them with Voyager 1 electron spectra for 2010. They also thank Mirko Boezio, Nico De Simone, Valeria Di Felice and their colleagues for proving PAMELA electron observations. The partial financial support of the South African National Research Foundation (NRF) and the Space Sciences Division of the South African National Space Agency (SANSA) is acknowledged. We thank the reviewers for helpful comments.